\documentclass[10pt,letterpaper,twocolumn]{article} 

\usepackage{ol2}
\usepackage[draft]{hyperref}
\usepackage{amsfonts}
\usepackage{amsmath}
\usepackage{amssymb}
\usepackage{amsthm}
\newcommand{\bsqr}{$\vcenter{\hbox{\tiny$\blacksquare$}}$}
\begin{document}

\twocolumn[ 

\title{Discrete solitons in coupled active lasing cavities}


\author{Jaroslaw E. Prilepsky,$^{1,*}$ Alexey V. Yulin$^2$, Magnus Johansson$^{3}$, and Stanislav A. Derevyanko$^1$}

\address{
$^1$Nonlinearity and Complexity Research Group, Aston University,
 Aston Triangle, B4 7ET, Birmingham, UK
\\
$^2$Centro de Fisica Teorica e Computacional,  Universidade de Lisboa, Lisboa 1649-003, Portugal \\
$^3$Department of Physics, Chemistry and Biology (IFM), Link{\"o}ping University, SE-581 83 Link\"{o}ping, Sweden \\
$^*$Corresponding author: 	y.prylepskiy1@aston.ac.uk
}

\begin{abstract}
We examine the existence and stability of discrete spatial solitons in coupled nonlinear lasing
cavities (waveguide resonators), addressing the case of active defocusing media, where the gain
exceeds damping in the low-amplitude limit. A new family of stable localized structures is found:
these are bright and grey cavity solitons representing the connections between homogeneous and
inhomogeneous states. Solitons of this type can be controlled by the discrete diffraction and are
stable when the bistability of homogenous states is absent.
\end{abstract}

\ocis{190.1450, 190.4390, 190.4420}

 ] 

\noindent Due to the huge progress in semiconductor-based photonic devices, a multitude of
experimentally achievable concepts have arisen, aimed at performing optical processing and light
reconfiguration \cite{aa08,afo09}. One of these concepts relates to light manipulation in
semiconductor microcavities closed by high-reflectivity Bragg reflectors
(microresonators). The cavity medium consists typically of quantum well structures which can be
absorbing or (with electrical or optical pumping) provide gain \cite{afo09}. The soliton
solutions in driven optical cavities (cavity solitons, CS) have
attracted much attention because of their potential applications in information
processing and optical memory schemes: The ability to control their switch-on/off process by an
address beam, and their location by introducing gradients in the
holding beam makes them interesting as mobile pixels for reconfigurable arrays of all-optical
processing units \cite{afo09}. The CS are a particular realization of the dissipative soliton
notion -- localized structures existing due to the balance between dissipation,
nonlinearity and diffraction \cite{aa08}.

A relatively new direction in the light processing in optical cavities refers to coupled systems of microresonators
\cite{lsc08,pel04,els07,ycs08,yc10,epl05,elk07} where the excitation level of each
cavity (or of an ordered pattern) can serve as an elementary 'pixel'. These
'pixels' are based on yet another type of dissipative solitary structures  -- the discrete cavity
solitons (DCS), which are localized excitations in coupled arrays of nonlinear cavities.
Achieving controlled manipulations with 'pixels' requires a detailed study of the properties
of these objects: nucleation thresholds, stability, mobility etc
\cite{lsc08}. It has to be noted that, unlike the continuum case \cite{afo09}, the studies of DCS
have so far only dealt with the case of a passive cavity, i.e. the losses were assumed to dominate
over the lasing gain \cite{pel04,els07,ycs08,yc10,epl05,elk07}. At the same time there is an interesting idea
of self-sustained light spots in cavities with active gain \cite{afo09}, where no
external driving is needed for supporting the stability of the CS (pixels). Therefore additional
efforts are needed to address specifically the case of DCS in \textit{active} media.

In this Letter we, for the first time to our knowledge, consider the
properties of DCS in coupled \textit{active cavities} and demonstrate that their structure and
stability can be drastically different from their counterparts in passive systems
\cite{lsc08,pel04,els07,ycs08,yc10,epl05,elk07}. In addition, we show that by adjusting the coupling (which is an example of a
\textit{discrete diffraction management}) one can control the stability of the DCS (a similar idea
for passive planar resonators was used in \cite{sek08}) which can be implemented as a useful tool
for all-optical processing in discrete systems.


We consider a regular array of evanescently coupled identical nonlinear cavities with
high-reflectivity mirrors placed at their facets, driven by a homogeneous holding beam having a
normal incidence. The mean-field model for the light distribution within the
array can be derived from
the coupled-mode equations \cite{pel04} assuming that i) the frequency detuning from the cavity
resonance is small (high-finesse cavities) and ii) the temporal dynamics is slow compared to the
round-trip time, i.e. the structure is short compared to the coupling and nonlinearity lengths (an
effective cavity length is on the order of 1-2$\mu$m, though much longer structures can also be
fabricated \cite{afo09}). Then the system of (normalized) equations governing the dynamics of
the averaged beam amplitude inside the $n$-th cavity, $A_n$, reads as:
\begin{eqnarray}\label{eq} \Big(\! \! \!
\! \! \! &i& \! \! \! \! \! \! \! \frac{d}{dt} + \Delta +  \alpha |A_n|^2 \Big)A_n + \nonumber \\\! \! &+& \! \!\! \! \!  \, C\big(A_{n+1}+A_{n-1}-2 A_n
\big) - i F(A_n)
 = P,
\end{eqnarray}
where all coefficients are real, $C>0$ characterizes the coupling strength, $\tilde{\Delta} =
\Delta - 2 C$ is the detuning of the pump frequency from the resonant one, $\alpha$ characterizes
the strength of the nonlinear (Kerr) polarization, $P>0$ is the amplitude of the holding beam and
$F(A_n)$ describes dissipative effects. The field $A_n$ is normalized to the lasing saturation
intensity $I_0$: $A_n/\sqrt{I_0} \to A_n$, and the time to the cavity round trip time
(see also \cite{pel04}).

\begin{figure}[tb]
\centerline{\includegraphics[width=7.5cm]{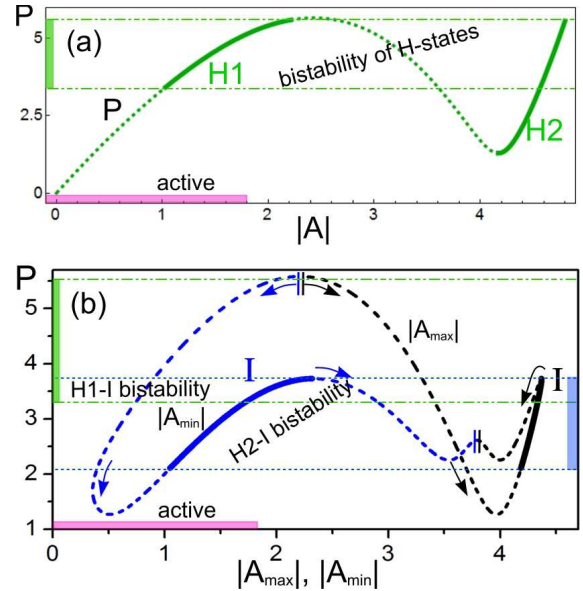}}
\caption{(Color online.) Response curves for (a) homogeneous (H) and (b) 2-site periodic (I)
solutions, with stable
regions highlighted by thick solid lines. (a) $P(|A|)$ for H-solutions, with stable lower
(H1) and higher (H2) states. (b) $P(|A_{max}|)$ and $P(|A_{min}|)$ for the
I-solution; the direction of `evolution' from the curve coalescence points (marked by $||$)
is identified with
arrows. Regions of H-H and H-I bistability are marked on the $P$-axes; the active region
($F(|A|)>0$) is highlighted on the $|A|$-axes.}
\label{fig1}
\end{figure}

The existence and stability properties of DCS in
model (\ref{eq}) can be essentially different from those of the continuous counterparts
\cite{pel04,lsc08}. In this Letter we will specifically concentrate on \textit{highly discrete
solutions} occurring in the case of defocusing nonlinearity, $\alpha<0$,
existing up to a limiting value of $C$ and having no continuum limit.
We opt for a simple but physically important and general form of dissipative function
with gain saturation \cite{lsc08,yc10,kkr08}: $F(A_n) = A_n [ \gamma (1 + |A_n|^2)^{-1} - \delta]$,
and the active region corresponds to amplitudes satisfying $F(|A|)>0$ (highlighted part of the
horizontal axis in Figs.\ref{fig1},~\ref{fig3}). Here parameters $\delta,\gamma>0$ describe linear losses and gain
respectively. We will assume that in the low-power limit the system is \textit{active}, so that
$\gamma>\delta$, distinguishing the case considered here from other studies of DCS
\cite{lsc08,pel04,els07,ycs08,yc10,kkr08,epl05,elk07}. This implies that in the absence of a
holding beam ($P=0$)
the trivial solution $A_n \equiv 0$ of Eq.(\ref{eq}) is unstable.

To specify the order of coefficients entering
Eq.(\ref{eq}) we took parameters for GaAs semiconductor lasers \cite{aa08} with the linear
refractive index $n_0=3.55$, saturation intensity $I_0=10^8\, \mathrm{W/cm^2}$, the Kerr
coefficient $n_2=-2~\times~10^{-13}~\mathrm{cm^2/W}$, the linear gain 1.5cm$^{-1}$ and the loss 0.5~cm$^{-1}$. Assuming the cross-section of resonators being $5 \mu$m$^2$, we
get that the unit of the normalized amplitude corresponds to the power $0.5$mW. The round trip
frequency is $\Omega \sim 10~^{10} s^{-1}$. With this in mind we set all dimensionless coefficients
$\Delta$, $\delta$, $\alpha$ and $\gamma$ in order of unity, which appears to be achievable in
experiments. For the visualization of our results we opt for the following particular values:
$\Delta=3.5$, $\delta=0.4$, $\gamma=1.7$, $\alpha = - 0.2$, yielding $|A|<1.8$ as active region.
For distances $\sim 1 \div 10$$\mu$m the coupling strength varies in a large
interval, so we can use a relatively weak realistic coupling $C=0.15$, a value which for the
chosen set of parameters is
slightly below the upper existence limit for the solutions found below.

\begin{figure}[t]
\centerline{\includegraphics[width=7.7cm]{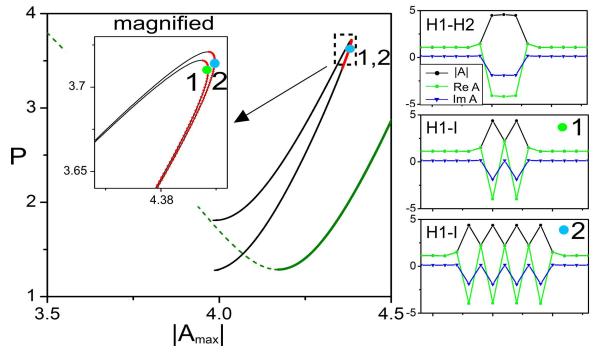}}
\caption{(Color online.) Left pane: Snaking diagram for bright DCS corresponding to H1-I-H1 connection
(stable regions are highlighted). The inset shows a magnified part of the snake. Right panes show
particular field distributions for DCS: the top pane shows a stable DCS corresponding to the H1-H2-H1
connection ($P=4.3$), the two lower panes show the distributions for the stable DCS corresponding to
H1-I-H1. The latter correspond to the points marked $\bullet$  '1', '2' on the snake.}\label{fig2}
\end{figure}

\begin{figure}[t]
\centerline{\includegraphics[width=6cm]{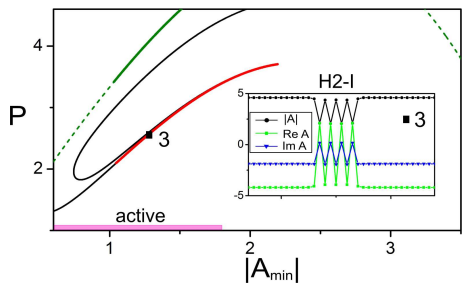}}
\caption{(Color online.)  Snaking diagram $P(|A_{min}|)$ for the grey DCS corresponding to the H2-I-H2 connection.
The inset shows the particular distribution of stable DCS referring to the point marked as \bsqr~'3'.}\label{fig3}
\end{figure}

The analysis of DCS in the system (\ref{eq}) starts from the identification of
stable homogeneous (H) states that serve as a background for localized solutions
\cite{lsc08,pel04,els07,ycs08,yc10}: setting $A_n=A$ (or $C=0$) in Eq.(\ref{eq}) one finds the
so-called response curve $P(|A|)$, plotted in Fig.\ref{fig1}(a) for the parameters chosen. The
static state is stable when Eq.(\ref{eq}) linearized above it has no time-growing solutions.
When two stable H-states coexist (thick solid parts marked H1 and H2 in
Fig.\ref{fig1}(a)), solitons for nonzero $C$ can generally be found as
\textit{connections} between these states \cite{yc10}: the H1-H2-H1 connection for bright solitons,
where the core intensity exceeds the background, or the H2-H1-H2 connection for grey solitons
with a 'dent' in the H2 `substrate'. In our case, as for passive
systems \cite{lsc08,pel04,els07,ycs08,yc10}, such stable DCS exist in a certain interval
within the
bistability region of H-states caused by the Kerr nonlinearity; 
an example of a stable bright soliton is
given in Fig.\ref{fig2} (upper right pane).
The major difference of H-states in the active case is the instability of the low-amplitude
part of the H1-branch, so that, in contrast to passive systems
\cite{lsc08,pel04,els07,ycs08,yc10}, DCS formed by such H-states
connections {\em cannot be stable at low $P$ values}.

Now we turn to another possible background for the formation of DCS -- \textit{inhomogeneous
periodic static state} (we label it as an I-state). In general, infinitely many extended solutions
with inhomogeneous patterning can be found but here, to demonstrate the main features
introduced by coupling to the I-states, we deal with the simplest family having a 2-site
period. Setting in Eq.(\ref{eq}) $A_{n}, \, A_{n+2}, \, \ldots = A_{max}$, and $A_{n+1}, \,
A_{n+3}, \, \ldots = A_{min}$, $A_{max} \neq A_{min}$, we solve the system for these two complex
fields. The results are given in Fig.\ref{fig1}(b) in the form of two (equivalent) response curves:
$P(|A_{max}|)$ and $P(|A_{min}|)$ (stable regions are marked with a thick solid line). The
I-background is discrete and does not exist in the continuum limit $C \to \infty$. One can observe
that the regions of stability for H-states and I-state partially overlap. This means that one can
compose new families of localized solutions in the form of a fragment of the I-state embedded in a
stable H-background (or vice versa). In addition,
\textit{the I-state can be stable below the region of H-bistability},
i.e. for parameters where one cannot have stable DCS in the form of H-states
connections. Thus, in this region the only stable DCS represent connections of type H2-I.

In Figs.\ \ref{fig2} and \ref{fig3} we show examples of
the {\em new type of DCS corresponding to H-I connections} found in our model.
We present them on snaking diagrams where each point
corresponds to a particular DCS. The bright/grey DCS are characterized by the maximal/minimal
amplitude of the soliton so the snakes corresponding to each soliton type are plotted on the curves
$P(|A_{max}|)$, (Fig.\ref{fig2}, corresponding to H1-I-H1 connection) and $P(|A_{min}|)$,
(Fig.\ref{fig3}, H2-I-H2 connection). These DCS are discrete, i.e. they can exist only up to a
finite value of coupling.
They have a nonuniform central excited part corresponding to a fragment of the I-state embedded
into either H1 or H2 background.  Importantly, the grey DCS of this type can be stable in the
region where no solitons representing the connections of H-states can be stable (see
Fig.\ref{fig3}), so this stability is achieved solely by the discrete diffraction. The solitons outside the stable regions typically display nonoscillatory exponential growth of small perturbations.

We note that, to the best of our knowledge, the DCS with H-I connections of the type found above
have not been studied in works on (passive) coupled cavity systems, although a solution
involving connections with a more complicated I-state was mentioned in
\cite{yc10}. Solutions involving oscillatory patterns at the tails were discussed in \cite{epl05,elk07}: in \cite{epl05} DCS with oscillatory decaying tails were found, while in \cite{elk07} the staggering low-amplitude plane-wave served as a background I-state for a grey DCS. Notably, both these types of DCS could be stable out of the H-bistability region. Our solutions have the staggering in the middle and come from the anticontinuous limit with generally large staggering oscillations. By decreasing $\gamma$ we checked that stable analogues of the
H-I type DCS with a 2-periodic I-state in fact do exist {\em also for passive systems}.

Our findings have several important physical consequences. First, we have demonstrated that in
active cavities, aside from 'traditional' DCS representing connections between H-states, there
can be stable solitary solutions with a lower power corresponding to the H-I connections,
Figs.\ref{fig2}, \ref{fig3}. Thus exerting an address beam for the nucleation of a 'pixel' one can
end up with an H-I rather than an H-H soliton. In turn, the mobility characteristics \cite{elk07} of H-I solitons
are likely to be different from those of the well-studied H1-H2 solutions, and the conditions for
optical processing can significantly deviate from those of the 'conventional' H-H DCS. Second, by adjusting
the discrete diffraction (i.e. coupling) one can gain stable solitary solutions in the region of
parameters, where no stable H1-H2 DCS can exist. Such stability control via a simple discrete diffraction
management can be used as an additional tool for light manipulation: one can control not only
the stability, but also the geometry of DCS providing more versatility for optical processing.
Finally, the only stable type of DCS suitable for 'pixels' manipulations in active cavities at low $P$ values (or for a relatively high gain)
corresponds to the H-I solitons.

A.Y. and M.J. thank NCRG, Aston University, for kind hospitality. M.J. also acknowledges support
from the Swedish Research Council. A.Y. was supported by the FCT (Portugal), grant
PTDC/FIS/112624/2009.

\pagebreak

\section*{Informational Fourth Page}
In this section, please provide full versions of citations to
assist reviewers and editors (OL publishes a short form of
citations) or any other information that would aid the peer-review
process.

\end{document}